\documentstyle[aas2pp4]{article}

\def\simless{\mathbin{\lower 3pt\hbox
     {$\rlap{\raise 5pt\hbox{$\char'074$}}\mathchar"7218$}}}   
\def\simmore{\mathbin{\lower 3pt\hbox
     {$\rlap{\raise 5pt\hbox{$\char'076$}}\mathchar"7218$}}}   
\def\msun{{\rm M}_\odot}                                       

\received{}
\accepted{}
\slugcomment{Submitted to Ap.J.Letters., June 1998}

\lefthead{M. M\'endez et al.}
\righthead{Kilohertz QPO peak in 4U 1608--52}

\begin{document}

\title{Kilohertz Quasi-Periodic Oscillation Peak Separation is
not Constant in the Atoll Source 4U 1608--52}

\author{M. M\'endez\altaffilmark{1,2},
        M. van der Klis\altaffilmark{1,3},
        R. Wijnands\altaffilmark{1},
        E. C. Ford\altaffilmark{1},
        J. van Paradijs\altaffilmark{1,4},
        B. A. Vaughan\altaffilmark{5}
}

\altaffiltext{1}{Astronomical Institute ``Anton Pannekoek'',
       University of Amsterdam and Center for High-Energy Astrophysics,
       Kruislaan 403, NL-1098 SJ Amsterdam, the Netherlands}

\altaffiltext{2}{Facultad de Ciencias Astron\'omicas y Geof\'{\i}sicas, 
       Universidad Nacional de La Plata, Paseo del Bosque S/N, 
       1900 La Plata, Argentina}

\altaffiltext{3}{Department of Astronomy, University of California,
Berkeley, Berkeley, CA 94720, USA}

\altaffiltext{4}{Physics Department, University of Alabama in Huntsville,
       Huntsville, AL 35899, USA}

\altaffiltext{5}{Space Radiation Laboratory, California Institute of
       Technology, MC 220-47, Pasadena CA 91125, USA}

\begin{abstract}

We present new Rossi X-ray Timing Explorer observations of the low-mass
X-ray binary 4U 1608--52 during the decay of its 1998 outburst.  We
detect by a direct FFT method the existence of a second kilohertz
quasi-periodic oscillation (kHz QPO) in its power density spectrum,
previously only seen by means of the sensitivity-enhancing `shift and
add' technique.  This result confirms that 4U 1608--52 is a twin kHz QPO
source.  The frequency separation between these two QPO decreased
significantly, from $325.5 \pm 3.4$ Hz to $225.3 \pm 12.0$ Hz, as the
frequency of the lower kHz QPO increased from 470 Hz to 865 Hz, in
contradiction with a simple beat-frequency interpretation.  This change
in the peak separation of the kHz QPOs is closely similar to that
previously seen in Sco X--1, but takes place at a ten times lower
average luminosity.  We discuss this result within the framework of
models that have been proposed for kHz QPO.  Beat frequency models where
the peak separation is identified with the neutron star spin rate, as
well as the explanations previously proposed to account for the similar
behavior of the QPOs in Sco X--1, are strongly challenged by this
result.

\end{abstract}


\section{Introduction}

In the past two years the Rossi X-ray Timing Explorer (RXTE) has
discovered kilohertz quasi-pe\-ri\-od\-ic oscillations (kHz QPOs) in eighteen
low-mass X-ray binaries (LMXB; see van der Klis \cite{vanderklis98} for
a review).  In almost all cases the power density spectra of these
sources show twin kHz peaks that move up and down in frequency together
as a function of mass accretion rate, keeping a separation consistent
with being constant (e.g., Strohmayer et al.  \cite{strohmayer96a};
Wijnands et al.  \cite{wijnandsetal97}; Ford et al.  \cite{ford97a}).
In some sources a third QPO peak has been detected during type-I X-ray
bursts, at a frequency consistent with the frequency separation of the
twin peaks (Strohmayer et al.  \cite{strohmayer96a}) or twice that value
(Smith, Morgan, \& Bradt \cite{smith97}; Wijnands \& van der Klis
\cite{wijnands97}; Zhang et al.  \cite{zhang96}; Wijnands et al.
\cite{wijnandsetal97}), indicating a beat-frequency interpretation.  The
third peak, while not strictly constant in frequency but varying by up
to 0.4\,\%, has been interpreted as being close to the neutron star spin
frequency (Strohmayer et al.  \cite{strohmayer96a}).

Until recently, Sco X--1 stood out as the only example where the
separation between the two simultaneous kHz QPOs was not constant, but
varied by 40\,\% (van der Klis et al.  \cite{vanderklisetal97a}) as the
mass accretion rate increased, posing a serious challenge to the simple
beat-frequency interpretation.  Using data of the 1996 outburst, and a
new technique to increase the sensitivity to weak QPOs, M\'endez et al.
(\cite{mendez98}) found the second of the twin peaks in 4U 1608--52, and
presented evidence for similar variations by 26\,\% of the peak
separation in this source, although only significant at the $3.5 \sigma$
level.

In this paper we report on the results from recent RXTE observations
obtained during the decay of the 1998 outburst of 4U 1608--52.  We
observed again two QPOs, and show conclusively that the peak separation
in 4U 1608--52 changes by more than 40\,\%.  The changes are remarkably
similar to those seen in Sco X--1.

\section{Observations}

We used the proportional counter array (PCA) onboard RXTE to observe
4U 1608--52 between February 6 and April 25 1998.  The observations were
triggered at the peak of the outburst, as measured by the RXTE All Sky
Monitor (ASM) experiment, and were planned to sample the source over the
whole decay of the outburst.  The data set consists of 40 observations,
each of them with exposures between 600 s and 19,980 s, and a total
observing time of 191 ks.  Figure \ref{asm} shows the ASM light curve
during the outburst, where we indicate the dates when we carried out the
pointed PCA observations.

Beside the two Standard modes that are available to all RXTE-PCA
observations, we collected data using additional modes with high time,
and moderate energy resolution, covering the nominal $2-60$ keV energy
band of the RXTE-PCA.  We also included a set of burst trigger and
catcher modes to record burst data at high time and energy resolution.
We recorded 3 type-I X-ray bursts.  A timing analysis of these data
revealed no burst oscillation, with 95\,\% confidence upper limits to
the amplitude of 6\,\% rms at the peak of the bursts, and 20\,\% at the
beginning or the end.

\section{Data Analysis and Results}

We used the Standard 2 data and the standard PCA background model (Stark
et al.  \cite{stark97}) for the spectral analysis.  First we divided the
data into 4 energy bands, $2.0 - 3.5 - 6.4 - 9.7 - 16$ keV, and
constructed a color-color and hardness-intensity diagram.  Based on the
spectral data, and on the power density spectra (described below) in the
range $0.01 - 128$ Hz, we conclude that 4U 1608--52 slowly moved from
the upper banana, through the lower banana to the island state (Hasinger
\& van der Klis \cite{hasinger89}) as the $2-60$ keV count rate decayed
from $\sim 14,500$ c/s to $\sim 260$ c/s.

We also produced background subtracted X-ray spectra that were well
fitted by means of a model consisting of a blackbody and a power law
with a high-energy cut-off.  Assuming a distance of 3.6 kpc (Nakamura et
al.  \cite{nakamura89}), during our observations the $2-20$ keV
luminosity of 4U 1608--52 decayed from $4.8 \times 10^{37}$ erg s$^{-1}$
at the highest count rate, to $1.2 \times 10^{36}$ erg s$^{-1}$ at the
lowest count rate.  Details of the spectral fits, color-color diagrams
and the analysis of the low frequency part of the power spectra will be
presented elsewhere.

\begin{figure}[ht]
\plotfiddle{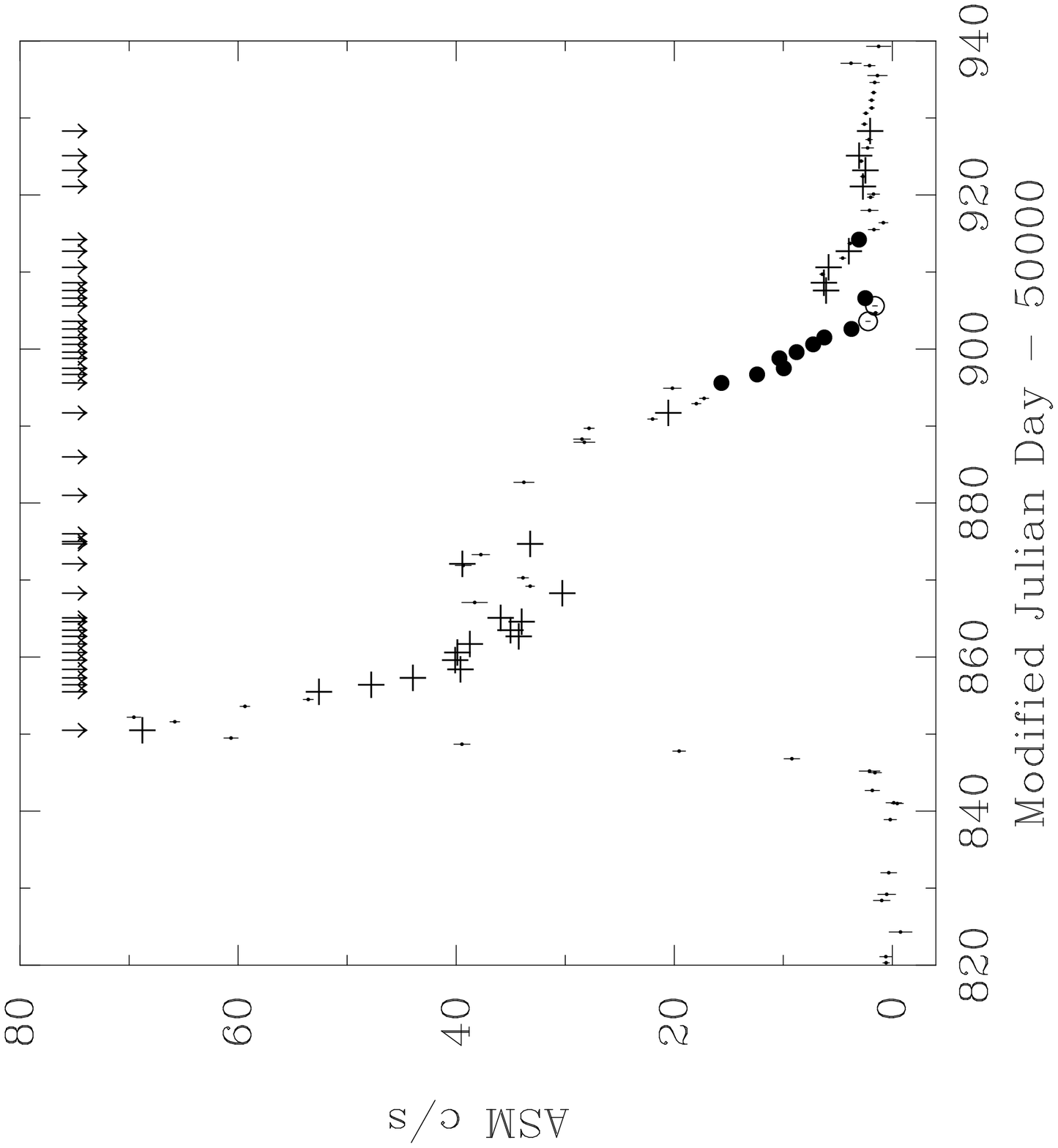}{250pt}{270}{70}{70}{-270}{250}
\vspace{6cm}
\caption{
ASM light curve of the 1998 outburst of 4U 1608--52.  The arrows
indicate the dates when we performed the pointed RXTE observations.
Filled circles indicate the observations where we detect, directly
or with shifts, 2 simultaneous kHz QPOs. Open circles and crosses
indicate the observations where we detect 1 and no kHz QPOs,
respectively.
\label{asm}
}
\end{figure}

\begin{figure}[ht]
\plotfiddle{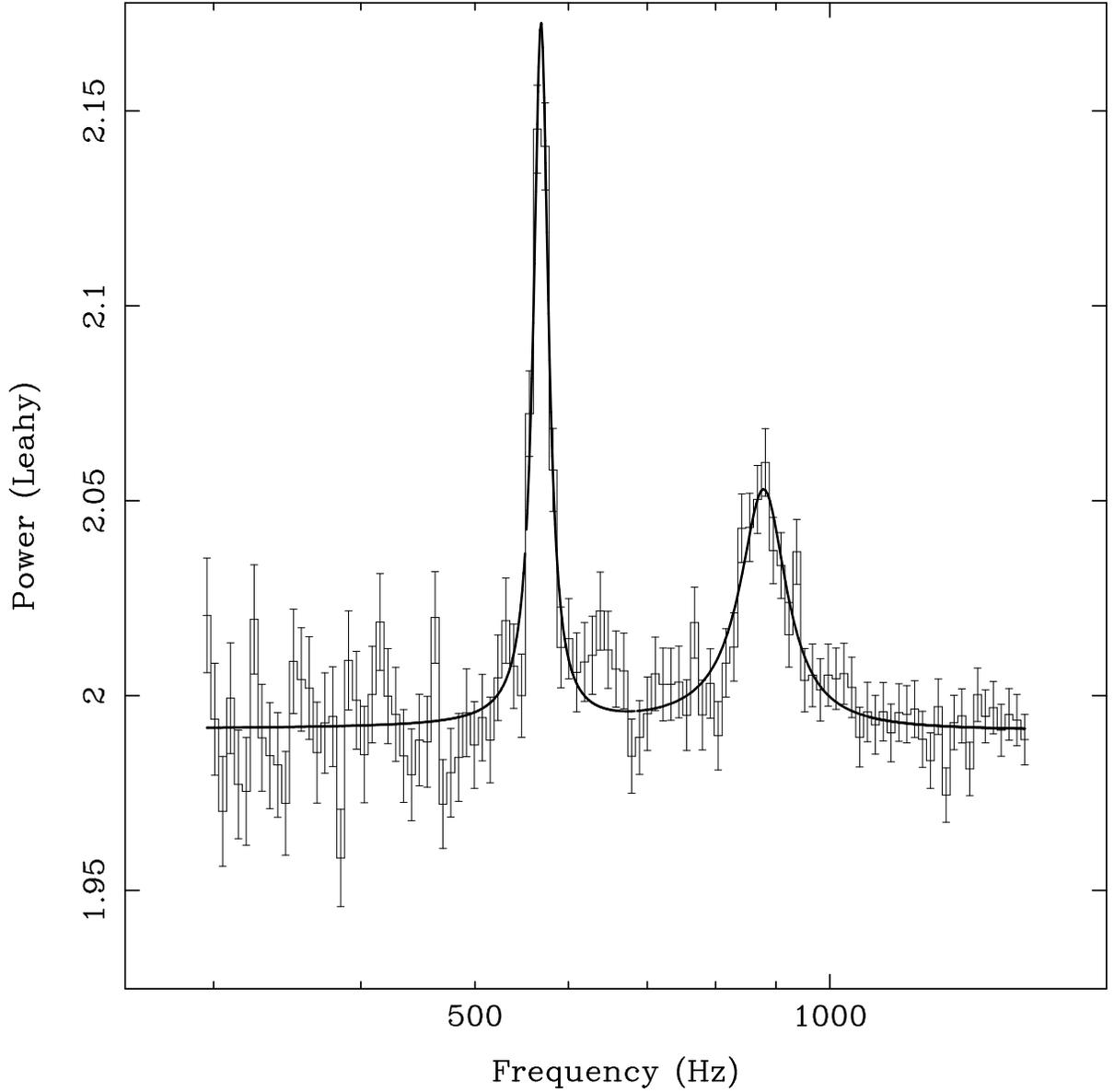}{220pt}{0}{90}{90}{-240}{-330}
\vspace{7.4cm}
\caption{
Power spectrum of a segment of 4224 s starting on UTC 1998 March 25
16:16, for the full energy band of the PCA.  Both kHz QPOs are visible.
No shift was applied to this power spectrum (see text).  On 64 s time
scales the FWHM of the lower QPO was $\sim 5$ Hz, but due to the
variation of its central frequency during the observation it appears
broader in the average power spectrum.  The upper QPO was broader, and
its FWHM on each segment of 64 s was $\sim 100$ Hz.
\label{figps}
}
\end{figure}

We divided the high-time resolution data into segments of 64 s, and
calculated a power spectrum for each segment preserving the maximum
available Nyquist frequency.  For each of the 40 observations we
produced an average power spectrum that we searched for QPOs at
frequencies above $\simmore 100$ Hz.  None of the observations obtained
before March 23 showed any QPO in this frequency range.  The 95\,\%
confidence upper limits on the amplitude for a peak with a FWHM of 10,
50, and 100 Hz were 0.8\,\%, 1.2\,\%, and 1.6\,\% rms, respectively on
February 6, and 2.0\,\%, 3.8\,\%, and 4.6\,\% rms, respectively on March
19.  Starting on March 23, when the luminosity of the source ($2 - 20$
keV) had dropped to $1 \times 10^{37}$ erg s$^{-1}$, we detected QPO
peaks in 12 observations (Fig.  \ref{asm}), at frequencies between 400
and 1000 Hz.  In 5 observations, this straight Fourier analysis revealed
two simultaneous QPOs, while in the remaining 7 observations we only
detected one QPO peak.  In Fig.  \ref{figps} we show an example of an
observation where we detected two simultaneous QPOs at $\sim 550$ Hz
(hereafter the lower QPO), and at $\sim 900$ Hz (hereafter the upper
QPO).

For those 7 observations where we only detected a single peak in the
average power spectrum we applied the same technique we described in
M\'endez et al.  (\cite{mendez98}).  We fitted the central frequency of
the QPO in each segment of 64 s (or longer when occasionally we had to
average some of the 64 s segments to detect the QPO), and then shifted
the frequency scale of each spectrum to a frame of reference where the
position of the peak was constant in time.  Finally, we averaged these
shifted power spectra.  This procedure revealed a second QPO in 5 of
those 7 observations, in 3 occasions at higher frequencies, and in 2
observations at lower frequencies than those of the peak originally used
to align the individual power spectra.  For the 2 remaining observations
(March 31 and April 5) we did not detect a second peak.  The 95\,\%
confidence upper limits on the amplitude of a second (undetected) QPO
with a FWHM of 10 Hz, 50 Hz, and 100 Hz were 3.5\,\%, 4.7\,\%, and
5.3\,\% rms, respectively on March 31, and 5.2\,\%, 8.1\,\%, and 10\,\%
rms, respectively on April 5.

For the observation of March 25, when the two QPOs were more or less
equally strong, we measured their FWHM in three energy bands, $3.5 - 6.8
- 11.2 - 14.9$ keV; the width of the lower and upper peak were $8.3 \pm
0.4$ Hz, and $88 \pm 12$ Hz, with no significant dependence on energy.

In the rest of this paper we only present the results of the analysis of
those data where, either directly or with shifts, we detected two
simultaneous kHz QPOs in the power spectrum.  These are 10 observations
of the 1998 outburst (Fig.  \ref{asm}, filled circles), plus 2
observations of the 1996 outburst (March 3 and 6; M\'endez et al.
\cite{mendez98}), yielding a total of 1,336 power spectra each of them
of 64 s of data.  During these observations the $2 - 60$ keV fractional
amplitudes of the lower and upper QPO varied from 5.3\,\% to 9.1\,\%
rms, and from 3.3\,\% to 8.8\,\% rms, respectively, and the FWHM varied
from 4.3 Hz to 9.4 Hz, and from 53 Hz to 173 Hz, respectively.

\begin{figure}[ht]
\plotfiddle{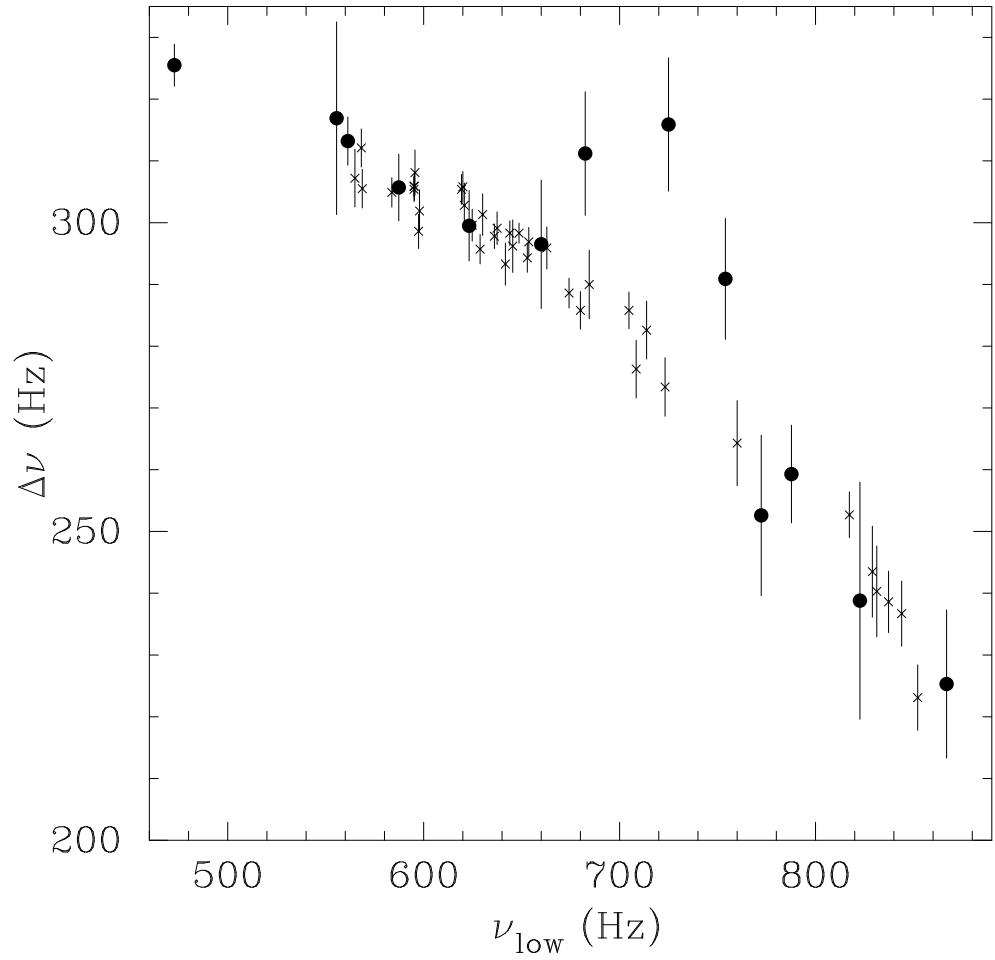}{250pt}{270}{70}{70}{-270}{250}
\vspace{6cm}
\caption{
The frequency separation between the upper and the lower QPO peak as a
function of the frequency of the lower QPO for 4U 1608--52 (circles).
The crosses are the data obtained for Sco X--1 (van der Klis et al.
\cite{vanderklisetal97a}). The rise of $\Delta \nu$ at $\nu_{\rm low}
\sim 700$ Hz in 4U 1608--52 is marginally significant.
\label{figdif}
}
\end{figure}

We aligned all the 1,336 power spectra using the lower QPO as a
reference, and we grouped the data in 13 sets of $\sim 50$ to $\sim 150$
power spectra, such that the frequency of the QPO did not vary by more
than $\sim 10-20$ Hz within each set.  Finally, we combined these
aligned spectra to produce an average power spectrum for each set. We
fitted all these 13 power spectra in the range $256 - 3000$ Hz using a
function consisting of a constant, representing the Poisson noise, and
two Lorentzians, representing the QPOs.  The fits were good, with
reduced $\chi^{2} \leq 1.1$, and the significance of both peaks was
always $> 3\sigma$.  In Figure \ref{figdif} we plot the frequency
difference, $\Delta \nu$, between the upper and the lower QPO as a
function of the centroid frequency of the lower QPO, $\nu_{\rm low}$:
As $\nu_{\rm low}$ increases from $\sim 475$ Hz to $\sim 865$ Hz, the
separation gradually changes from $325.5 \pm 3.4$ Hz to $225.3 \pm 12.0$
Hz, i.e.  by $100.2 \pm 12.5$ Hz.  In Figure \ref{figdif} we also
plotted $\Delta \nu$ versus $\nu_{\rm low}$ as measured for Sco X--1
(van der Klis et al.  \cite{vanderklisetal97a}).  Despite the big
difference in luminosity, $\sim L_{\rm Edd}$ for Sco X--1 and $\sim 0.1
\times L_{\rm Edd}$ for 4U 1608--52, both sources behave in the same
manner.

In order to get a significant measurement of the properties of the upper
QPO as a function of the frequency of the lower QPO, we had to average
together power spectra from intervals that were very distant in time.
To check whether this way of averaging the data affected our results we
selected three contiguous intervals for which the frequency of the lower
QPO remained approximately constant, allowing us to measure reasonably
well both QPOs simultaneously.  During 2624 s, 2624 s, and 3264 s of the
observations of 1998 March 26 and 27, and 1996 March 3 the lower QPO
remained more or less constant at $\sim 600$ Hz, $\sim 770$ Hz, and
$\sim 870$ Hz, respectively.  In the worst case (1996 March 3) the
significance of the detection of the upper peak was $2.7 \sigma$.  These
three intervals showed exactly the same trend as seen in Figure
\ref{figdif}, only the error bars were larger.  For these three
intervals $\Delta \nu$ was $298 \pm 9$ Hz, $277 \pm 23$ Hz, and $231 \pm
13$ Hz, respectively.

\section{Discussion}

We have confirmed the existence of the second kHz QPO in 4U 1608--52,
and we have conclusively shown that $\Delta \nu$, the frequency
separation of both peaks, is not constant, but changes as a function of
the frequency of the lower kHz QPO.  This is the first case where this
is observed to occur in an atoll source, at a luminosity that is far
below the Eddington luminosity.  A similar result was obtained
previously by van der Klis et al.  (\cite{vanderklisetal97a}) only for
Sco X--1, a Z-source at a near-Eddington mass accretion rate.

Three different models have been proposed to explain the kHz QPOs.  In
the photon bubble oscillations model (Klein et al.  1996a,b) the flow of
matter is funneled onto the polar caps of a highly magnetized neutron
star accreting at high rates, and the radiation pressure becomes locally
super-Eddington.  The accretion column develops radiation-hydrodynamic
turbulence, and energy is transported to the surface of the accretion
column by photon bubbles, which produce oscillations in the luminosity.
Klein et al.  (\cite{klein96b}) applied this model to explain the 20,
40, and 60 Hz QPO and the overall shape of the power spectrum observed
in the high-luminosity binary pulsar GRO J1744--28 (Fishman et al.
\cite{fishman95}; Kouveliotou et al.  \cite{kouveliotou96}) which
contains a neutron star with a high magnetic field ($B \sim 2.4 \times
10^{11}$ G; Cui \cite{cui97}), and to the kHz QPO observed in the
luminous Z-source Sco X--1 (van der Klis et al.  \cite{vanderklis96}),
which is accreting matter at a high rate (Hasinger \& van der Klis
\cite{hasinger89}).  It is widely accepted that in the atoll sources the
magnetic field is $\simless 10^{8-9}$ G.  In 4U 1608--52 during these
observations the X-ray luminosity was less than 10\,\% of the Eddington
luminosity for a $1.4 \msun$ neutron star.  Combining these numbers, and
using standard magnetospheric accretion theory, the area of the polecap
where the matter accreted onto the neutron star was between 20 and
60\,\% of the total stellar surface (see Frank, King, \& Raine
\cite{franketal92}, eq.  [6.14]).  It is therefore quite unlikely that
the accretion rate was locally super-Eddington.  This argues, at least
for this source, against the photon bubble interpretation of the kHz
QPOs.

Titarchuk \& Muslimov (\cite{titarchuk97}), and Titarchuk, La\-pi\-dus, \&
Muslimov (\cite{titarchuk98}) proposed that all the QPOs observed in Z
and atoll sources might be explained in terms of the rotational
splitting of oscillation modes of a nearly Keplerian accretion disk.
One of the predictions of this model is that the ratio of the frequency
of the upper to the lower kHz QPO, $\nu_{\rm upp} / \nu_{\rm low}$, is
only a function of $H/R$, the ratio of the half-thickness of the disk to
the radius of the annulus in the disk at which the oscillations occur.
The ratio $\nu_{\rm upp} / \nu_{\rm low}$ (Titarchuk et al.
\cite{titarchuk98}, eq.  [A9] for $m = -2$ and $k=1$) can vary from 5
(for $H \ll R$) to 1.37 (for $H \sim R$).  When we detected kHz QPOs,
the luminosity of 4U 1608--52 was $\simless 10$\,\% $L_{\rm Edd}$.  For
these luminosities, and for any realistic structure of the disk, one
expects $H \ll R$.  During our observations of 4U 1608--52, $\nu_{\rm
upp} / \nu_{\rm low}$ gradually decreased from $1.69 \pm 0.01$ at
$\nu_{\rm low} = 475$ Hz to a value of $1.26 \pm 0.01$ at $\nu_{\rm low}
= 875$ Hz, which are inconsistent with the predicted value of 5.  In Sco
X--1, despite the large difference in luminosity, and therefore large
difference of the inferred structure in the inner part of the disk,
$\nu_{\rm upp} / \nu_{\rm low}$ changes in the same manner, varying from
$1.54 \pm 0.01$ at 565 Hz to $1.26 \pm 0.01$ at 852 Hz (van der Klis et
al \cite{vanderklisetal97a}).  Moreover, in both cases $\nu_{\rm upp} /
\nu_{\rm low}$ reaches lower values than those allowed by the model.
These results show that the disk oscillation model as presented cannot
account for the observed kHz QPOs in these two sources.

The most widely accepted models for the kHz QPOs are beat-frequency
models where the upper kHz QPO, $\nu_{\rm upp}$, represents the
Keplerian frequency of the accreting material in orbit around the
neutron star at some preferred radius (van der Klis et al.
\cite{vanderklis96}), while the lower frequency peak, $\nu_{\rm low}$,
is produced by the beating of $\nu_{\rm upp}$ with another frequency,
$\nu_{\rm S}$, identified as the spin frequency of the neutron star.
Strohmayer et al.  (\cite{strohmayer96b}) proposed that $\nu_{\rm upp}$
is produced at the magnetospheric radius, while Miller et al.
(\cite{miller98}) proposed that it originates at the sonic radius.  As
$\nu_{\rm S} = \nu_{\rm upp} - \nu_{\rm low}$, these models predict that
the frequency difference $\Delta \nu$ of the twin kHz peaks should
remain constant, although $\nu_{\rm upp}$ and $\nu_{\rm low}$ may vary
in time.  This kind of models provides a natural explanation for the
fact that in most of the sources where two simultaneous kHz QPOs have
been observed $\Delta \nu$ does not change significantly as the
frequency of the two QPOs varies, while they are also consistent with
the presence of a third kHz peak that is sometimes detected during
type-I bursts (Strohmayer et al.  \cite{strohmayer96a}; Smith, Morgan,
\& Bradt \cite{smith97}) at a frequency near $\nu_{\rm S}$ or $2
\nu_{\rm S}$.

Two explanations have been put forward to account for the variable peak
separation in Sco X--1, both of them related to the inferred
near-Eddington mass accretion rate in this source.  White \& Zhang
(\cite{white97}) proposed that the photosphere of the neutron star might
expand by 35\,\% while conserving its angular momentum, and therefore
slows down as $\dot M$ increases.  Alternatively, at near-Eddington
accretion the height of the inner disk might increase and the change of
$\nu_{\rm upp} - \nu_{\rm low}$ might reflect different values of
$\nu_{\rm upp}$ at different heights in the disk (Lamb, private
communication).

However, we know now that in the low luminosity source 4U 1608--52
$\Delta \nu$ also changes significantly, in contradiction with a
beat-frequency interpretation involving the neutron star spin.  Neither
of the above explanations can account for this result in 4U 1608--52, as
it is accreting matter at rates far below the Eddington critical rate.
One might argue that 4U 1608--52 and Sco X--1 are different, in some
respect, from the rest of the sources that show kHz QPOs.  But if this
were the case, it would be difficult to explain why the properties of
the kHz QPOs observed in all sources (including both 4U 1608--52 and Sco
X--1) are so homogeneous.  Moreover, Psaltis et al.  (\cite{psaltis98})
recently showed that for nine other sources where the two QPOs have been
observed simultaneously, $\Delta \nu$ is consistent both with being
constant and with having a similar behavior as that observed in Sco X--1
and 4U 1608--52.

In the beat-frequency model the lower kHz peak is expected to be at
least as broad as the upper peak, as it is generated by a beat between
the upper QPO and the (coherent) neutron star spin.  This is contrary to
what we observe for 4U 1608--52, where the upper peak is intrinsically
broader than the lower peak.  For instance, during 3840 s on 1998 March
25, when the frequency of the lower peak varied between 550 Hz and 575
Hz, the FWHM of the lower and upper peak were $9.3 \pm 0.8$ Hz, and
$95.4 \pm 13.8$ Hz ($2-60$ keV), respectively.  A rapidly variable
scattering medium around the neutron star could cause the broadening,
and might even be responsible for the frequency shift, if it has a
(quasi-) periodicity of its own (Lamb, private communication).  The QPO
at $\nu_{\rm upp}$, which in the sonic model (Miller et al.
\cite{miller98}) is a beaming oscillation, would be more sensitive to
this than the peak at $\nu_{\rm low}$, which is supposed to be produced
by oscillations of the luminosity.  If this interpretation is correct,
it also needs to explain why the width of the upper QPO does not depend
on energy.

At this point, none of the modified beat-frequency models proposed can
explain the varying peak separation in 4U 1608--52 (and in Sco X--1).
In order to maintain a beat-frequency model, the change in $\nu_{\rm
upp} - \nu_{\rm low}$ by $\sim 40$\,\%, both in 4U 1608--52 and
Sco X--1, requires a similar change in $\nu_{\rm S}$.  The spin of the
neutron star cannot change by such a large factor on such a short
timescale.  It might still be possible to rescue the beat-frequency
interpretation, although at the expense of its simplicity.  Perhaps the
idea of a layer at the surface of the neutron star that does not
corotate with the body of the star can still be applied to explain these
results.  The detection of coherent oscillations during type-I burst in
4U 1608--52 might provide a strong clue to the nature of the kHz QPOs in
LMXBs, as this would at least make clear at which mass accretion level,
if any, the kHz peak separation does approach the inferred spin rate.

\acknowledgements

We are thanked to Prof.  Hale Bradt for his comments that helped us to
improve the original manuscript.  This work was supported in part by the
Netherlands Organization for Scientific Research (NWO) under grant PGS
78-277 and by the Netherlands Foundation for research in astronomy
(ASTRON) under grant 781-76-017.  MM is a fellow of the Consejo Nacional
de Investigaciones Cient\'{\i}ficas y T\'ecnicas de la Rep\'ublica
Argentina.  JVP acknowledges support from the National Aeronautics and
Space Administration through contract NAG 5-3269 and 5-4482.  MK
gratefully acknowledges the Visiting Miller Professor Program of the
Miller Institute for Basic Research in Science (UCB).

\onecolumn
\clearpage

\clearpage

\clearpage

\end{document}